\documentclass[fleqn,twoside]{article}
\usepackage{espcrc2}
\usepackage{amsmath}
\usepackage{amssymb}

\usepackage{graphicx}

\def\lsim{\mathrel{\rlap{\lower4pt\hbox{\hskip1pt$\sim$}}
    \raise1pt\hbox{$<$}}}         
\def\gsim{\mathrel{\rlap{\lower4pt\hbox{\hskip1pt$\sim$}}
    \raise1pt\hbox{$>$}}}         

\newcommand{\kkbar}{K^0-\overline{ K^0}}
\newcommand{\bbar}{B^0-\overline{ B^0}}
\newcommand{\bbbar}{B^0_{s(d)}-\overline{ B^0}_{s(d)}}

\newcommand{\bbbard}{B^0_d-\overline B^0_d}
\newcommand{\bea}{\begin{eqnarray}}
\newcommand{\eea}{\end{eqnarray}}
\newcommand{\beq}{\begin{equation}}
\newcommand{\eeq}{\end{equation}}
\newcommand{\gev}{{\rm GeV}}
\newcommand{\mev}{{\rm MeV}}
\newcommand{\msbar}{{\overline{\rm MS}}}

\title{Lattice results relevant to the CKM matrix determination}
\author{Damir Becirevic\address[]{Dipartimento di Fisica, 
        Universit\`a di Roma ``La Sapienza", Piazzale Aldo Moro 2, I-00185 Rome,
	Italy}%
        }
       
\begin{document}

\begin{abstract}
A brief discussion of the recent lattice studies 
of the quantities that are relevant to the Standard Model 
description of CP-violation in the hadronic sector is presented. 
A comment on $B\to K^\ast \gamma$ decay is made too.
\vspace{1pc}
\end{abstract}

\maketitle

\section{Introduction}
The most popular way to estimate the amount of CP-violation in the Standard Model 
is to study the shape of the CKM unitarity triangle  
involving the third generation of quark flavors. The standard 
unitarity triangle analysis (UTA)~\cite{uta}, which requires the least theoretical 
assumption, comprises the $\kkbar$ and $\bbbard$ mixing amplitudes, 
as well as the SU(3) breaking ratio in the $\bbar$ system. 
Additional constraint comes from information on $\vert V_{ub}/V_{cb} \vert$ ratio,  
which will be discussed towards the end of this review.
The three ``golden" mixing amplitudes over-constrain the CKM-triangle through 
a hyperbola and circles in the $(\bar \rho, \bar \eta)$-plane which emerge from 
the comparison of the theoretical predictions with the experimental
determinations of $\left|\varepsilon_K \right|$, $\Delta m_d$ and the limit  
 on $\Delta m_s/\Delta m_d$. Besides the CKM parameters, the theoretical expressions 
contain the Wilson coefficients and the hadronic matrix elements.
The Wilson coefficients encode the information about physics at high energy scales, they  
are computed in perturbation theory and are known at NLO accuracy 
(for a review see e.g.~\cite{buras}). 
The hadronic matrix elements, instead, describe the low energy  QCD dynamics and their 
computation requires a theoretical control of non-perturbative QCD. To be specific, 
the following matrix elements need to be computed:
\bea \label{me}
\hspace*{-2mm}\langle \bar K^0 \vert O_{sd}(\mu) \vert  \bar K^0\rangle &=& {8\over 3} m_K^2 f_K^2
 B_K(\mu)\,,\hfill\nonumber \\ 
\hspace*{-2mm}\langle \bar B_d^0 \vert O_{bd}(\mu) \vert  \bar B_d^0\rangle &=& {8\over 3} m_{B_d}^2 f_{B_d}^2
 B_{B_d}(\mu)\,,\hfill \\ 
\hspace*{-2mm}\langle \bar B_s^0 \vert O_{bs}(\mu) \vert  \bar B_s^0\rangle &=&
 {8\over 3} m_{B_s}^2 f_{B_s}^2
 B_{B_s}(\mu)\,, \hfill \nonumber
\eea
where the operator $O_{Qq} = (\bar Q q)_{V-A}(\bar Q q)_{V-A}$. On the r.h.s. of eq.~\eqref{me} we recognize 
the pseudoscalar decay constants $f_K$ and $f_{B_{d(s)}}$, and the familiar ``bag" parameters, 
$B_K(\mu)$ and $B_{B_{d(s)}}(\mu)$. That is where the lattice QCD enters the stage and 
provides us with a completely non-perturbative way to calculate these (and other) hadronic quantities. 
The following two points should be emphasized:~\footnote{For details about formulating QCD 
on the lattice, please see e.g.~\cite{reviews}. }
\begin{itemize}
\item[$\otimes$] Lattice QCD is not a model but the approach allowing the computation of 
various Green functions from first principles of QCD: the only 
parameters entering the computation are those that appear in the QCD lagrangian, 
namely the quark masses and the (bare) strong coupling;
\item[$\otimes$] QCD is solved numerically on the lattice by using 
the Monte-Carlo methods which induce the statistical errors in the results. 
The central limit theorem applies and ensures that those errors fall as $1/\sqrt{N_{\rm conf}}$, 
with $N_{\rm conf}$ being a number of independent SU(3) gauge field 
configurations on which a given Green function has been calculated. 
Thus, in principle, computation of the hadronic quantities on the 
lattice can be made to an arbitrary accuracy.
\end{itemize} 
In practice, however, various approximations have to be made, each 
introducing some systematic uncertainty into the final results. Yet all those 
approximations are improvable by: (a) increasing the computing power 
so that, one by one, all the approximations can be removed from the computation; 
(b) improving the approach theoretically so that the hadronic quantities, 
computed on the lattice, converge faster to their values in the continuum 
limit.

Now I return to the lattice determination of the ``golden" hadronic 
quantities entering the standard UTA. The present situation is discussed at length in 
ref.~\cite{lellouch}, and the actual results are:
\bea \label{wa's}
&&\hat B_K = 0.86(6)(14) \,,\nonumber\\
&&\hat B_{B_d} =1.34(12) \,,\nonumber \\
&& f_{B_d} = 203(27)\left(^{+00}_{-20}\right)\ \mev \,,\\
&&f_{B_s}/f_{B_d} = 1.18(4)\left(^{+12}_{-00}\right)  \,,\nonumber \\
&&\xi =(f_{B_s}\sqrt{B_{B_s}})/(f_{B_d}\sqrt{B_{B_d}}) = 1.18(4)\left(^{+12}_{-00}\right)\;.\nonumber
\eea
The above quantities are completely dominated by the systematic uncertainties, some of  
which are discussed in what follows.
 
\section{$\kkbar$ mixing} 

\begin{figure}[t]
  \includegraphics*[width=\columnwidth]{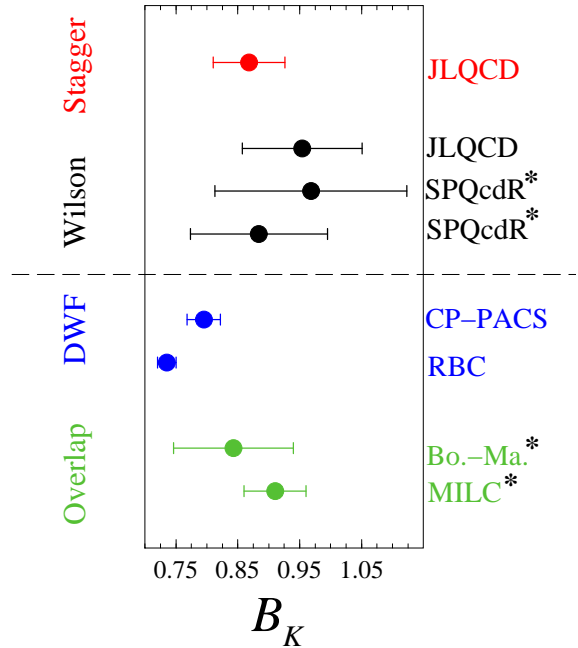}
\vspace*{-10mm}
\caption{\footnotesize  Quenched lattice results for the $\hat B_K$ parameter from 
4 different lattice actions. The values above/below the dashed horizontal line are 
obtained after/without extrapolating to the continuum limit. The second result by SPQcdR 
is obtained by using the method proposed in ref.~\cite{no-subtraction}.} 
\label{figBk}
\end{figure}

Results of the precision lattice computations are presented as $B_K^\msbar(2\ \gev)$, 
in the quenched approximation, i.e. the quark loops in the background gauge field 
configurations are being neglected, $n_f=0$. In the real physical world, however, one  
needs $n_f=3$. Numerically, the conversion from the $\msbar$(NDR) scheme to the renormalization 
group invariant (RGI) form, $\hat B_K$, is almost 
completely independent of $n_f$ (see discussion in ref.~\cite{qcd02}). 
By using $\Lambda_\msbar^{n_f=3} =338(40)$~MeV~\cite{bethke}, I obtain~\footnote{For 
a comparison, with $\Lambda_\msbar^{n_f=0} =250(25)$~MeV, one has 
$\hat B_K = 1.388(8) \cdot B_K^\msbar (2\ \gev)$.}
\bea
\hat B_K = 1.382(15) \cdot  B_K^\msbar (2\ \gev)\,. 
\eea 
A map of the precision quenched lattice QCD determinations of 
$\hat B_K$ is shown above the dashed line in fig.~\ref{figBk}, 
where the asterisks denote the results presented this year. 
The most accurate is the value obtained by using the so-called {\it staggered} 
fermions~\cite{jlqcd-stagger}, $\hat B_K = 0.87(6)$.  The advantage of that action 
is that it preserves the chiral symmetry, but the drawback is that it breaks the flavor symmetry. 
With Wilson fermions, instead, the flavor symmetry is preserved but the chiral symmetry 
is sacrificed.  The loss of the chirality induces the additive renormalization of the 
operator $O_{sd}$, which is more involved than the (usual) multiplicative one. 
The non-perturbative method to subtract these mixing has been first implemented in 
ref.~\cite{jlqcd-wilson} by using the quark Ward identities, and then in 
ref.~\cite{spqcdr-wilson} by using the so-called RI/MOM method~\cite{Roma}. 
The problem of spurious mixing has been alleviated by a judicious application 
of the Ward identity on the parity violating operator which does not suffer 
from the spurious mixing problem on the lattice~\cite{no-subtraction}. That 
proposal has been implemented in the high statistics lattice simulation and 
the result, in the continuum limit, reads  $\hat B_K = 0.88(11)$.  
The results below the horizontal line in fig.~\ref{figBk} do not refer to 
the continuum limit. Their major significance lies in the fact that they are 
obtained by using the actions that satisfy both the chiral and the flavor 
symmetry. Computations with these actions are much more involved and require 
huge computational resources. This year, first results for $\hat B_K$, obtained 
by using the so-called overlap fermions, have been reported by the MILC~\cite{milc-BK} 
and by the Boston-Marseille~\cite{giusti} collaborations. Even though it is early 
to draw conclusions, the observed good agreement with the values above the horizontal 
line in fig.~\ref{figBk} is encouraging. Somewhat lower are the values obtained by 
using the so-called domain wall fermions (DWF). Although their statistical errors 
are very small, in my opinion, their systematic errors are not fully realistic 
(see discussion in ref.~\cite{qcd02}). Nevertheless, it should be stressed  
that the results for $\hat B_K$ obtained by using DWF agree --within the error 
bars-- with other lattice determinations. The remaining systematics is almost entirely 
due to quenching. Although the first (low statistics) lattice studies suggest that 
the unquenching is not changing the value of $B_K$~\cite{unquenching-Bk}, we follow 
ref.~\cite{sharpe} and assume that the quenching error is of the order of $15 \%
$. Thus it is of upmost importance, and challenging task for the lattice community, 
\underline{to unquench $B_K$}.

\section{$\bbbar$ mixing}

In this case, besides the light quark, one also has to deal with the 
heavy quark on the lattice. Since the Compton wavelength of 
the realistic $b$-quark is smaller than the presently attainable lattice 
resolution, one has to adopt some other strategy in order to compute the 
properties of the heavy-light hadrons. Four ways to deal with that problem 
have been implemented in the quenched computation of the decay constant 
$f_{B_{s(d)}}$, and/or the bag-parameter $\hat B_{B_{s(d)}}$.\\
\indent
$\odot$ \underline{Lattice QCD}: 
one can afford to work with propagating heavy quarks of masses around the 
physical charm quark mass and then one has to {\it extrapolate} to the $b$-quark 
by means of standard heavy quark scaling laws~\cite{ff-symmetries};\\
\indent
$\odot$ \underline{Heavy quark effective theory} (HQET)  on the lattice 
is very difficult to extend beyond the static limit ($m_b\to \infty$). 
The long standing problem to renormalize the 
axial current non-perturbatively has been solved recently (see eg.~\cite{sommer}). 
The static results will therefore be possible to use in constraining the 
extrapolations of the results obtained in the full theory, i.e. with propagating 
heavy quarks;\\
\indent
$\odot$ \underline{Non-relativistic QCD} (NRQCD), is the approach in which the 
$1/m_b$-corrections are included in both the lagrangian and the operators. 
Besides difficulties in renormalizing the operators in such a theory on the 
lattice, a particularly problematic is the fact that the expansion is 
in $1/(a m_b)$  ($a$ being the lattice spacing), so that the continuum limit 
($a\to 0$) cannot be taken;\\
\indent
$\odot$ \underline{ Fermilab } (FNAL) approach consists in using the lattice QCD 
with propagating quarks and ``sending" the heavy quark mass over the lattice 
cut-off, followed by a redefinition of the hadron masses and reinterpretation of 
the theory in terms of a $1/m_b$-expansion. In some cases the approach is believed to be plagued by 
``renormalon shadows"~\cite{chris}.\\
\indent
None of the enumerated approaches is fully satisfactory on its own 
and, at present, they all should be used to check the consistency of the 
obtained results. For example, the results for $f_B$, as obtained by 
using various approaches, agree quite impressively~(see ref.~\cite{lellouch}). Nevertheless, 
the systematics ({\it within the quenched approximation}) cannot be further reduced 
until the fully QCD based lattice method is devised to compute the heavy-light quantities 
directly on the lattice. Otherwise, one should wait for the next generation of parallel 
computers to do the job. A candidate method for the precision computation of $f_B$ 
has been proposed this year in ref.~\cite{nazario}, in which the step scaling function 
has been employed to make a controllable $1/(L m_b)$-expansion, with $L$ being the side 
of the lattice box. From an exploratory study, $f_B=170(11)(23)$~MeV has 
been quoted. The authors promise to reduce the error bars very soon and to respond 
to the questions raised in ref.~\cite{yamada}.

The effect of unquenching has been extensively studied this year by the MILC 
collaboration~\cite{milc-nf=2}. They conclude that $f_B^{n_f=2}$ is about $15\%
$ larger than $f_B^{n_f=0}$. Moreover, in their preliminary study with $n_f=3$, 
they show that the ratio $f_{B_s}/f_{B_d}$ remains essentially unchanged 
when one switches from $n_f=0 \to 2 \to 3$~\cite{milc-nf=3}. 
That brings me to the important estimate of the hadronic parameter $\xi$ (see eq.~\eqref{wa's}). 
The SU(3) light flavor breaking effect in the bag parameter is equal to 
$B_{B_s}/B_{B_d}=1.00(4)$~\cite{lellouch,yamada}, and the whole effect of $\xi$ being different from 
unity comes from $f_{B_s}/f_{B_d}$. 
\begin{figure}[h!!]
\vspace*{-5mm}
  \includegraphics*[width=\columnwidth]{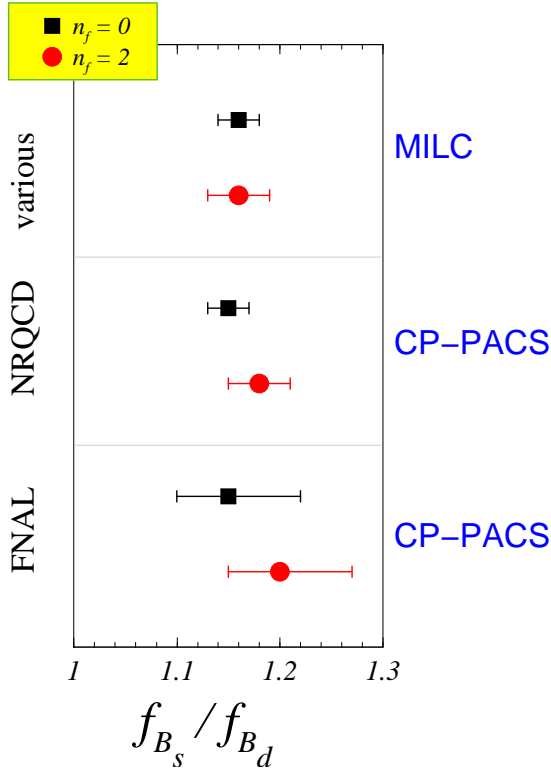}
\vspace*{-10mm}
\caption{\footnotesize  Quenched and unquenched ($n_f=2$) lattice estimates for the SU(3) 
breaking ratio $f_{B_s}/f_{B_d}$, as obtained by three different methods indicated on the left. 
For more details see refs.~\cite{milc-nf=2} (MILC) and~\cite{cp-pacs-nrqcd,cp-pacs-fnal} (CP-PACS).
No shift due to the chiral log term is displayed.} 
\label{fig:2}
\end{figure}

\subsection{Chiral logs in $f_{B_s}/f_{B_d}$}

Even if one gets a good handle over the heavy quark, there is still an extrapolation in 
the light quark mass to be made. With respect to the physical strange quark mass, the light quarks 
that are directly accessible from the lattice simulations lie in the range $1/2 \lesssim m_q/m_s
\lesssim 3/2$, in which one observes that the heavy-light decay 
constant changes nearly linearly under the variation of the light quark mass. 
In other words, $f_{B_s}$ is accessed directly, whereas an extrapolation in the light 
quark mass is necessary in order to reach $f_{B_d}$. It has been argued recently that if in 
the standard (linear) extrapolation one also includes the chiral logarithmic terms, 
the SU(3) breaking ratio may get shifted from $f_{B_s}/f_{B_d} = 1.16$, by $+0.16$, 
which is a $100\% 
$ error on the net SU(3) breaking~\cite{kronfeld}. In ref.~\cite{yamada} 
the shift can be as large as $+0.24$. There is an ambiguity while implementing the 
chiral logs in extrapolation. It is not clear at what mass scale the chiral 
logs dominate over the other, higher order, terms in the chiral expansion.
If one sets it for the light meson masses around $600$~MeV, the extrapolated value will 
be strongly shifted upwards (w.r.t. the result of linear extrapolation)~\cite{kronfeld}. 
The shift is roughly halved if the chiral logs are introduced for the  
light pseudoscalar mesons $\sim 350$~MeV. To avoid that ambiguity, ref.~\cite{fkfpi} 
proposes to study the double ratio
\bea \label{double}
R = {f_{B_s}/f_{B_d}\over  f_K/f_\pi}  \;,
\eea  
in which almost all the chiral log corrections cancel, and one can make the linear 
extrapolations with a small error. From the available lattice data with $n_f=2$, 
the NLO chiral correction in the ratio~\eqref{double} seems to cancel too. From that 
consideration, ref.~\cite{fkfpi} concludes that $f_{B_s}/f_{B_d} \simeq f_K/f_\pi$, 
where the latter ratio is known very well from the experiments, 
$f_K/f_\pi = 1.22(1)$. Therefore, the shift is indeed present but it is not as 
spectacular as previously thought. Similar conclusion has been reached by the MILC 
collaboration~\cite{milc-nf=2}.  They showed that if the chiral logarithms are 
consistently included in the entire analysis of the lattice data, the shift in 
$f_{B_s}/f_{B_d}$ is much smaller, namely $+0.04$. 

Notice that the second errors in $f_{B_d}$, $f_{B_s}/f_{B_d}$ and $\xi$, quoted in 
eq.~\eqref{wa's} (see also ref.~\cite{yamada}), reflect a guesstimate of the 
impact of the chiral logs on extrapolation in the light quark mass.
As we just explained, those errors cannot be too large, and I personally believe 
that through the strategies such as those proposed in refs.~\cite{milc-nf=2} 
and~\cite{fkfpi}, the lattice community will be able to cut down on those errors very soon.

\section{ $\vert V_{cb}\vert $ from $B\to D^{(\ast)} \ell \nu$ }
The extraction of $\vert V_{cb}\vert $ from the exclusive semileptonic modes, 
$B\to D^{(\ast)} \ell \nu$, requires the computation of the corresponding 
form factors. The cleanest method for their extraction has been proposed and implemented 
in ref.~\cite{Fermilab}.  It has been discussed at this conference by J.~Simone~\cite{jim}, 
and the reader is referred to those articles for more details. It would be very nice if  
the other lattice approaches adopt that strategy and compute the relevant form factors 
through the double ratios described in ref.~\cite{Fermilab}.

\section{ $\vert V_{ub}\vert $ from $B\to \pi (\rho) \ell \nu$ }

\begin{figure}[t!!]
  \hspace*{-5mm}\includegraphics*[width=1.1\columnwidth]{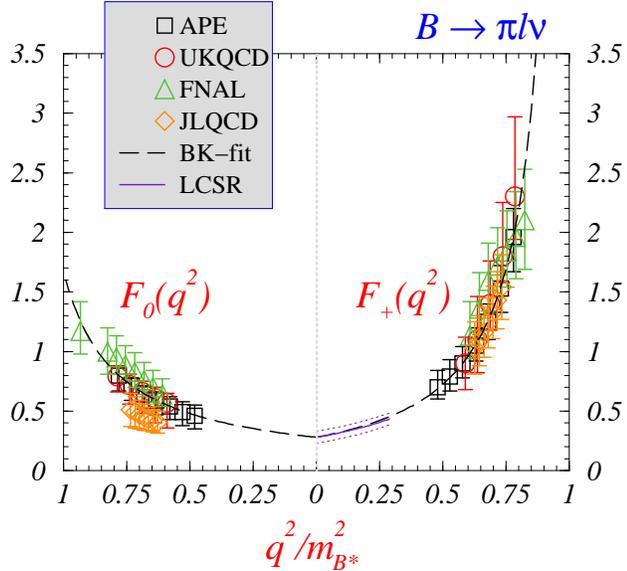}
\vspace*{-7mm}
\caption{\footnotesize Recent (quenched) lattice results for the two form factors are plotted 
head-to-head, since they satisfy the condition $F_+(0)=F_0(0)$. For illustration a fit to the BK-form~\eqref{bk}
is shown as well as the prediction obtained by using the light cone QCD sum rules (LCSR)~\cite{khodjamirian}.} 
\label{fig:3}
\vspace*{-5.5mm}
\end{figure}

In recent years several new lattice studies of the $B\to \pi$ form factors 
appeared~\cite{ukqcd-bpi,ape-bpi,fnal-bpi,jlqcd-bpi,shigemitsu-bpi}. 
The present situation with lattice results is depicted in fig.~\ref{fig:3}, 
where we plot the two form factors that parametrize the hadronic matrix element 
\bea
\langle \pi(p^\prime)\vert V_\mu \vert B(p)\rangle =c^+_\mu F_+(q^2) + c^0_\mu F_0(q^2)\,,
\eea
with $c_{+/0}$ being the known kinematic factors. The range of the transfer momenta available 
from this decay is huge: $0 \leq q^2 \leq (m_B-m_\pi)^2 = 26.4 \ \gev^2$. The lattice can 
be used only for large  $q^2$'s (smail recoil), and therefore the assumptions 
are necessary if we are to cover the entire $q^2$-range. In ref.~\cite{bpi-bk} a simple 
parametrization has been proposed which includes various symmetry constraints onto 
the shapes of the form factors~\cite{ff-symmetries,ff-symmetries2}, as 
well as the kinematic condition $F_+(0)=F_0(0)$. It is often referred to as 
BK and it reads
\bea \label{bk}
F_{+}(q^2) &=&  {c\   (1- \alpha) \over (1 - q^2/m_{B^\ast}^2) (1 - \alpha  q^2/m_{B^\ast}^2)} \,, \cr
F_{0}(q^2) &=&  {c \  (1- \alpha) \over 1 - q^2/(\beta m_{B^\ast}^2)} \,,
\eea
\begin{table}[hb!!]
\setlength{\tabcolsep}{0.3pc}
\caption{ \footnotesize  Results by various lattice groups are fitted to the 
form~\eqref{bk}. For comparison, the LCSR results are presented in the same 
form (last line). Note that instead of the parameter $c$, we quote $F(0)=c\ (1- \alpha)$.
\vspace*{1mm}}
\label{tab1}
\hspace*{-6.5mm}
\vspace*{-2mm}
\begin{tabular}{|ccc|c|}
\hline
\vspace*{-3.5mm}&&&\\
 $F(0)$ & $\alpha$ & $\beta$ &  ref. \\ \hline \hline
\vspace*{-3mm}&&&\\
 $0.30(^{+6}_{-5})(^{+4}_{-9})$ &   $0.46(^{+09}_{-10})(^{+37}_{-05})$ &  
 $1.27(^{+14}_{-11})(^{+04}_{-12})$ & {\small UKQCD~\cite{ukqcd-bpi}} \\
\vspace*{-3mm}&&&\\
 $0.28(6)(5)$ &   $0.45(17)(^{+06}_{-13})$ &  $1.20(13)(^{+15}_{-00})$ & {\small APE-I~\cite{ape-bpi}} \\
\vspace*{-3mm}&&&\\
 $0.26(5)(4)$ &   $0.40(15)(9)$ &  $1.22(14)(^{+15}_{-00})$ & {\small APE-II~\cite{ape-bpi}} \\
\vspace*{-3mm}&&&\\
 $0.33(^{+2}_{-3})$ &   $0.34(^{+9}_{-3})$ &  $1.31(^{+3}_{-9})$ & {\small FNAL~\cite{fnal-bpi} }\\
\vspace*{-3mm}&&&\\
 $0.23(^{+4}_{-3})$ &   $0.58(^{+12}_{-09})$ &  $1.28(^{+12}_{-20})$ & {\small JLQCD~\cite{jlqcd-bpi} }\\
 \hline 
\vspace*{-3mm}&&&\\
 $0.28(5)$ &   $0.32(^{+21}_{-07})$ &  -- & {\small KRWWY~\cite{khodjamirian}} \\
 \hline 
\end{tabular}
\end{table}
\noindent
thus consisting of three parameters only: $c$, 
$\alpha$ and $\beta$. Results of the fit of the lattice data to this form are given in 
table~\ref{tab1}. We notice a pleasant agreement with the results obtained by 
LCSR~\cite{khodjamirian}, which is believed to be the method of choice in the 
low $q^2$-region. We also see a good consistency of the results obtained by using three
different approaches. The agreement is especially good for the form factor $F_+(q^2)$, 
which is the only one entering the decay rate (for $\ell = e, \mu$). 
A bit less good is the agreement for the form factor $F_0(q^2)$.
It is worth stressing that so far all the computations are performed in the quenched 
approximation. Recently, in ref.~\cite{bpi-bpz}, it has been pointed out that the quenching 
effects may be significant, particularly for the form factor $F_0(q^2)$. That conclusion has 
been reached by confronting the corresponding expressions derived in the quenched and in the 
full ChPT. The remaining step, in my opinion, is to unquench the $B\to \pi$ form factors. 

In the case of $B\to \rho \ell \nu$, instead of $2$ one has $4$ form 
factors and less kinematic constraints, which makes the computation more 
complicated. Like in the case of $B\to \pi$ transition, the lattice results for $B\to \rho$
decay are accessible for relatively large values of $q^2$. The constrained 
extrapolations to the low $q^2$'s, guided by the symmetry relations~\cite{ff-symmetries,ff-symmetries2}, 
are however much more difficult to control in this case. To make the phenomenologically relevant predictions,  
the lattice results for $q^2 > 10\ \gev^2$ of ref.~\cite{spqr-brho} have been interpolated and then 
combined with the light cone QCD sum rule results of ref.~\cite{lcsr-bkstar}, which are again 
expected to be reliable in the region of low $q^2$'s. From that exercise and by using the 
experimental branching ratio $B(\bar B^0 \to \rho^+\ell \nu)$, recently measured by CLEO, BaBar 
and Belle~\cite{exp-brho}, we obtain
\bea
\vert V_{ub}\vert = 0.0034(6)\,.
\eea
The reader interested in new results for 
$B\to \rho \ell \nu$ decay form factors is invited to consult ref.~\cite{spqr-brho}. 

\section{ $B\to K^\ast \gamma$ }

\begin{figure}[t!!]
\includegraphics*[width=1.0\columnwidth]{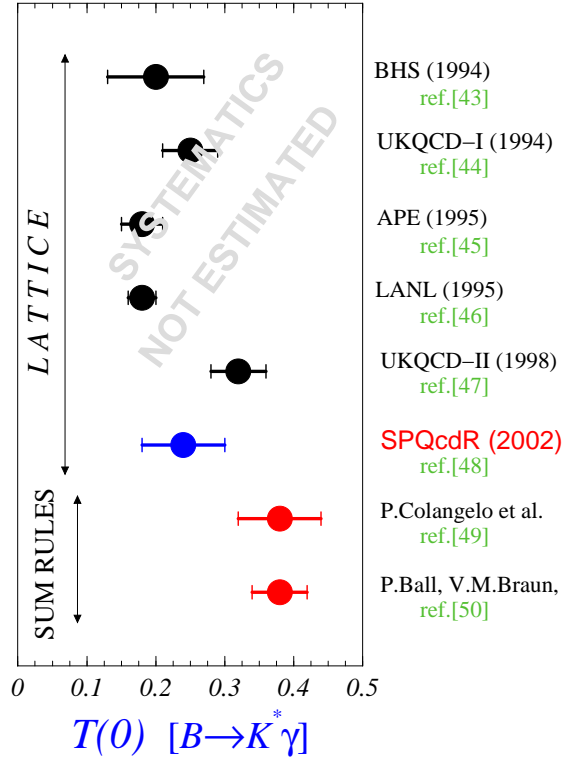}
\vspace*{-7mm}
\caption{\footnotesize Lattice results for the form factors $T^{B\to K^\ast}(0)$. Regarding the 
earlier lattice results, the values corresponding to  the ``pole/dipole" forms without the systematic 
uncertainty estimates are displayed. Also shown are the QCD sum rule estimates.} 
\label{compare-T}
\vspace*{-6mm}
\end{figure}
$b\to s \gamma$ is a fascinating decay mode: it is mediated by the flavor changing 
neutral current and therefore completely dominated by the loop effects. The hope is that by confronting 
theory Vs. experiment, one can probe the content in the loops and perhaps detect the non-Standard Model physics 
contributions. For that purpose the accurate Standard Model predictions are necessary. The NLO theoretical prediction 
for the inclusive $B\to X_s\gamma$ decay has been recently completed in ref.~\cite{misiak} (see also references therein). 
For the exclusive decays, instead, the situation is obscured by the necessity 
for the accurate computation of the hadronic form factors
\bea
\langle K^\ast(p^\prime)\vert \bar s \sigma_{\mu \nu}(1 + \gamma_5) b \vert B(p)\rangle =
\sum_{i=1}^3 c^{(i)}_{\mu \nu} T_i(q^2)\,,\nonumber
\eea
where the kinematic factors $c^{(i)}_{\mu \nu}$ are known (see e.g.~\cite{dima}). When the photon is on-shell, 
i.e. at $q^2=0$, $c^{(3)}_{\mu \nu}=0$ and  $T_1(0)=T_2(0)$. Previous lattice results were 
hampered by the lack of information about the heavy mass scaling behavior in the range of low $q^2$'s. 
That generated a proliferation of the $q^2$-forms to which the data have been fitted at 
large $q^2$'s, and then extrapolated to $q^2=0$. The needed extra information comes from 
the symmetry constraints discussed in refs.~\cite{ff-symmetries2,leet}, allowing one to write down a 
parameterisation similar to~\eqref{bk}. The ``pole/dipole" form discussed in 
refs.~\cite{bhs-bkstar,ukqcd-bkstar,ape-bkstar,lanl-bkstar} is also consistent with 
symmetry constraints~\cite{ff-symmetries2,leet}~\footnote{ 
The forms used in the reanalysis of the UKQCD data 
in ref.~\cite{ukqcd2-bkstar} are also consistent with the 
symmetry constraints discussed in~\cite{ff-symmetries2,leet}.}.
More details on the new lattice results by the SPQcdR collaboration can be found in ref.~\cite{spqr-bkstar}. 
The new (still preliminary!) result for $T_1(0)=T_2(0)\equiv T(0)$ is
\bea
T^{B\to K^\ast}(0) = 0.24(5)\left(^{+1}_{-2}\right)\;.
\eea
In fig.~\ref{compare-T} we compare the available results for the form factor $T(0)$ computed on the lattice, 
with the values obtained by using the QCD sum rules~\cite{qsr-bkstar,lcsr-bkstar}. We see that the 
lattice estimates are lower than the QCD sum rule ones. Although the QCD sum rule method is
more suitable for the kinematic corresponding to $q^2=0$, it is not clear where the 
discrepancy with the lattice QCD results comes from. More research in that direction is needed 
(as emphasized at this conference in ref.~\cite{ali}). In addition, it should be reiterated 
that so far all the lattice studies are made in the quenched approximation.

\section{Conclusion}

In this talk I discussed the status of the lattice QCD computation of the quantities that are 
relevant to the standard determination of the shape of the CKM unitarity triangle (see
ref.~\cite{stocchi} for the most recent update of the UTA). 
Quite a bit of progress has been made in gaining more control over the systematic 
uncertainties in the quenched lattice studies. A lot of work is still needed, especially 
in unquenching the current estimates. Besides the ``golden" quantities 
for the CKM triangle analysis, other quantities, such as the quark masses, decay constants, 
form factors, light cone wave functions,~\dots are  computed on the lattice. 
Accurate determinations of these quantities is very useful for theoretical studies of 
the non-leptonic heavy-to-light decays. 
Finally, I would like to mention the intensive activity, within the lattice 
community, in taming the $K\to \pi\pi$ decay amplitudes which I did not have 
space nor time to cover in this talk. For a status report please 
see~\cite{kpipi-1}. 

\vskip .5cm

\leftline{\bf Acknowledgments}

I am grateful to all my collaborators from the SPQcdR collaboration 
for their advice in preparing this talk. The correspondence with 
S.~Sharpe and N.~Yamada, many discussions with L.~Lellouch, as well as 
the help from C.~Maynard, T.~Onogi and J.~Simone in preparing the table~1, 
are kindly acknowledged. Finally, I thank the organizers for the 
invitation to this exciting and enjoyable conference.


\end{document}